\documentclass[preprint,12pt]{elsarticle}
\usepackage{amsmath}

\usepackage{graphicx}
\usepackage{longtable}

\usepackage{amssymb}

\biboptions{numbers,sort&compress}

\journal{Physica A}

\begin{document}

\begin{frontmatter}



\title{The ``S" Curve Relationship between Export Diversity and Economic Size of Countries}


\author{Lunchao Hu, Kailan Tian, Xin Wang and Jiang Zhang}

\address{Department of Systems Science, School of Management, Beijing Normal University, Beijing, China, 100875}
\ead{zhangjiang@bnu.edu.cn}

\begin{abstract}
The highly detailed international trade data among all countries in
the world during 1971-2000 shows that the kinds of export goods and
the logarithmic GDP (gross domestic production) of a country has an
S-shaped relationship. This indicates all countries can be divided
into three stages accordingly. First, the poor countries always
export very few kinds of products as we expect. Second, once the
economic size (GDP) of a country is beyond a threshold, its export
diversity may increase dramatically. However, this is not the case
for rich countries because a ceiling on the export diversity is
observed when their GDPs are higher than another threshold. This
pattern is very stable for different years although the concrete
parameters of the fitting sigmoid functions may change with time. In
addition, we also discussed other relationships such as import
diversity with respect to logarithmic GDP, diversity of exporters
with respect to the number of export goods etc., all of these
relationships show S-shaped or power law patterns. Although this
paper does not explain the origin of the S-shaped curve, it may
provide a basic empirical fact and insights for economic diversity.
\end{abstract}


\end{frontmatter}


\section{Introduction}
\label{sec.introduction}

Diversity phenomenon exist widely in economic systems. As we
observed, individual consumptions become more diversified when
people get richer\citep{anderson_long_2006}; organizations demand
for more diverse human resource, products and investments when they
become larger\citep{page_difference:_2007}; and countries can export
more kinds of products and services to other countries as their
economies are stronger. Nathan et.al. has discussed the relationship
between the network diversity and economic development by a large
data set of individual communication\citep{eagle_network_2010}.
Templet discovered that the economic diversity generally increases
in regional economy\citep{templet_energy_1999}. However, so far,
this kind of researches who addressed the diversity phenomena in
economics, especially the quantitative patterns based on empirical
data, are very sparse.

Nevertheless, compared to the situation in economics, there are many
studies on diversity phenomena in ecology both from the empirical
data and theoretical
models\citep{hubbell_unified_2001,gaston_biodiversity:_1998,mulder_physical_2001,loreau_biodiversity_2001}.
We can borrow some useful conceptions and methods from the ecology
to study the economic diversity because these two ``eco'' systems
have lots of similarities and
commonalities\citep{jerome_chave_scale_2003,nekola_wealth_2007}. For
example, the relationship between the number of species and the
area\citep{hubbell_unified_2001}, or the
productivity\cite{loreau_biodiversity_2001} of the habitat have been
widely discussed in ecology. This relationship inspired us to find
the similar relationship between the export diversity which is
compared to the species diversity and the economic size (GDP) of a
country which is compared to the area or productivity of a habitat
in international trade.

Although Krugman et al. have touched the problem of the relationship
between the number of different kinds of export goods (export
diversity) and the size of a country (population) in their classic
international trade
model\citep{krugman_increasing_1979,helpman_market_1987}, they
simply assumed that this relationship is linear without any support
of empirical
data\citep{armington_theory_1969,krugman_intraindustry_1981}.
Recently, Petersson has studied the export diversity and
intra-industry trade in Southern
Africa\citep{l_petersson_export_2005}. Johansson and Nilsson
reported {R\&D} accessibility and regional export diversity in
Sweden\citep{s._johansson_r&d_2007}. Although these studies were
based on the empirical data, they only focused on a specific country
or region but not the universal pattern on the global.

In this paper, we use the highly detailed international trade data
during 1971-2000 to investigate the relationship between the export
diversity of a country which is measured by counting the categories
of export goods and the size of an economy which is measured by its
GDP. Based on the data, we find that a nation with higher GDP level
will indeed export a wider variety of goods, however, this
dependence can be described by an S-shaped curve (fitted by a
logistic equation) which is one of the most important universal
patterns in complex
systems\citep{zwietering_modeling_1990,haanstra_use_1985,northam_urban_1979}
rather than the common linear or power law
relations\citep{hubbell_unified_2001}. Furthermore, we find that the
S-shaped curve is ubiquitous because it can be applied to the
relationships between other diversity-related variables such as the
number of different kinds of import goods, the number of exporters,
etc. and economic size. While the relationships between any pair of
the diversity-related variables can be described by power laws.

The outline of the paper is as follows. In section \ref{sec.data},
we briefly introduce the data source. Section \ref{sec.results}
proposes our main results. Subsection \ref{sec.three} discusses the
classification of countries into three stages. We show how robust
and stable the pattern is in subsection \ref{sec.alongwithtime}. And
also, we discuss how the key parameters in S-shaped curves indicate
the dynamical change of the global economy. In section
\ref{rec.discussion}, more relationships on diversity of
international economics are discussed. At last, section
\ref{sec.conclusion} points out that although this paper does not
provide any explanation of the S-shaped curves, it reveals a robust
and promising pattern which is the first step of the study on the
economic diversity phenomenon.

\section{Data}
\label{sec.data}

Our analysis is based on the world GDP statistics from the World
Bank web-site (www.worldbank.org) and the bilateral trade statistics
from NBER-UN world trade database (www.nber.org/data). The former
records the name of nearly 240 countries, as well as their GDP,
population, etc. from 1971 to 2006, while the latter records the
importer, exporter, 4-digit categories, the value and quantity of
each bilateral trade from 1962 to 2000\citep{hidalgo_product_2007}.

When dealing with the NBER-UN trade data, we calculated the total
number of the categories of export
commodities\citep{hubbell_unified_2001} for each country as the
measurement of the export diversity. We have also tried other
indices such as Shannon entropy to measure the export
diversity\citep{l_petersson_export_2005,s._johansson_r&d_2007}, but
they can't give universal and well-fitted patterns. The countries
without the export diversity or GDP data are ignored. Thus, we got
the full data of 126 countries from 1971 to 1991, 148 countries from
1992 to 2000.

The NBER-UN trade data were constructed from United Nations trade
flows over two periods: (i) 1962-1983, the data for which covered
all trade flows that different products were classified by SITC
Rev.1 standard (Standard International Trade Classification,
Revision 1) and trading partners with country codes; (ii) 1984-2000,
same as the first data set except the products were classified by
SITC Rev.2 standard.

It is necessary to mention some historical background of the
Standard International Trade Classification(SITC). SITC, Revision 1,
represented an improvement over the original SITC due to the
increase of the volume of trade and the diversity of commodity
during l960-l968 as well as the limitation of the SITC. And also, in
1984, the Economic and Social Council further replaced the SITC
Rev.1 by SITC Rev.2 because of the same reason. Therefore, we know
that it was the expansion of commodity diversity itself caused the
alteration of the classification standard but not the opposite.
Hence, our results are not dependent on the change of the
classification standard.

\section{Results}
\label{sec.results}

\subsection{S-shaped curve}
\label{sec.sshapedcurve}

\begin{figure}
\centerline {\includegraphics[width=18cm]{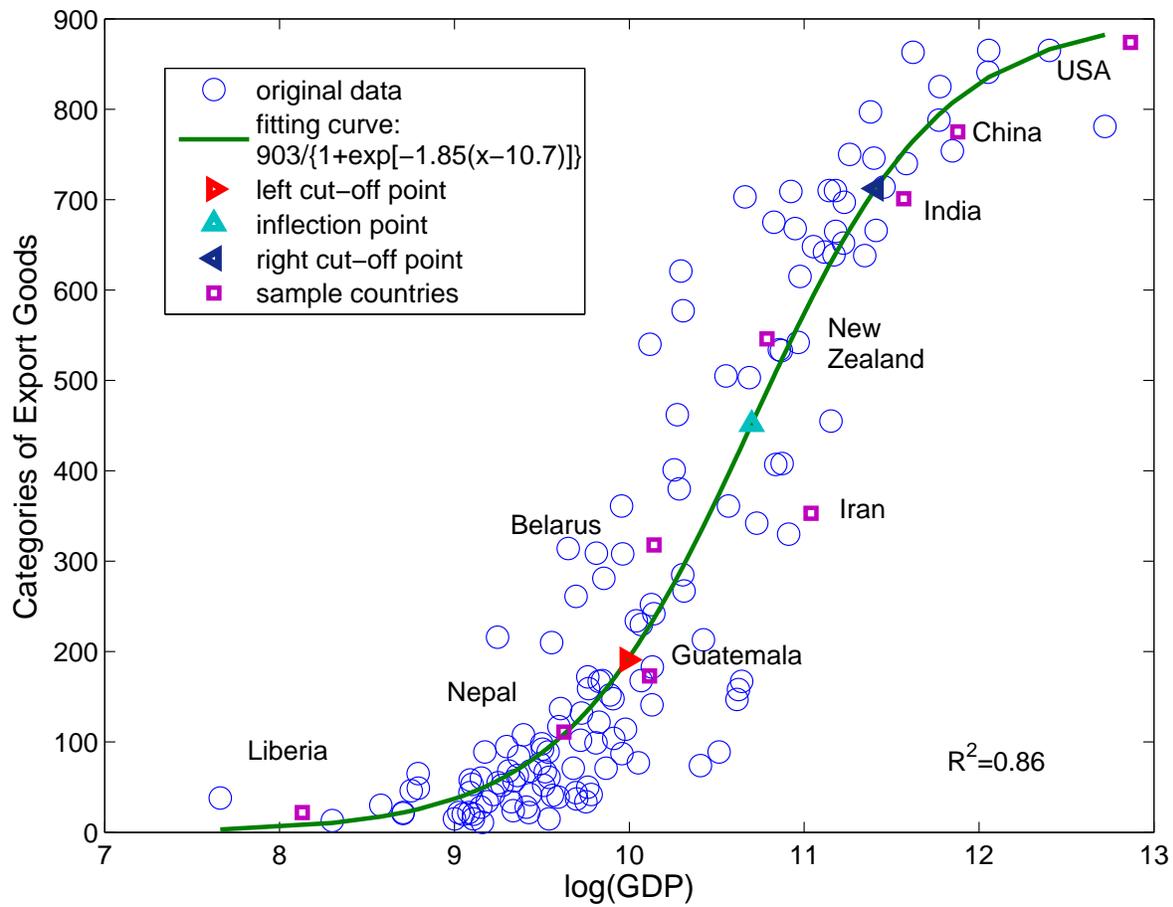}} \vskip3mm
\caption{The original data and the ``S'' curve fitting of export
diversity and log(GDP) for all countries in 1995. The curve has
three stages: initial stage, acceleration stage and final stage,
which are determined by the three critical points (left cut-off
point, inflection point, and right cut-off point) which are denoted
as triangles. Besides, some countries are picked out and are marked
by the squares.}\label{fig.diversity_gdp}
\end{figure}

As the gravity model of trade theory has found, the trade flows
(export/import volume or value) between any countries can be
described by a Newtonian gravity-like
equation\citep{krugman_international_2002,james_e._anderson_gravity_2011}.
That means the trade flows are the power law
functions\citep{zhang_allometric_2010} of the economic size of the
importer or exporter. However, the export diversity has a very
different dependence on GDP. Based on the collected data, we plotted
the original data and the estimated curve in Fig
\ref{fig.diversity_gdp}.

From the data shown in Fig \ref{fig.diversity_gdp}, we know that an
S-shaped curve may fit them very well. We use the logistic function
like the following Equation:
\begin{equation}\label{eqn.scurve}
y=\frac{A}{1+e^{-k(x-X_M)}},
\end{equation}
which are widely used in many
disciplines\citep{zwietering_modeling_1990} to characterize the
empirical S-shaped data. In equation \ref{eqn.scurve}, $y$ is the
export diversity measured by the number of different categories of
the export goods, and $x$ represents log(GDP). $A$, $X_M$ and $k$
are parameters to be estimated with the given data. $A$ is the upper
limit of the diversity of commodities. $k$ is the slope on the
inflection point and represents the maximum relative growth rate of
diversity. $X_M$ stands for the logarithmic GDP size at the
inflection point (see Fig \ref{fig.diversity_gdp}).

The S-shaped curve can be divided as three stages: in the first
stage, when a nation's economic size (GDP) stays in a low level, the
number of export goods is naturally small and increases with a
slower speed than the growth of GDP; in the second stage, their GDPs
climb up to a relatively high level, and the growth rate of the
diversity increases dramatically; in the last stage, the growth rate
of diversity slows down and gradually falls behind GDPs again.

By means of Least Square (LS) method, we obtained the estimated
values of parameters and some essential statistical data
(R-square=0.86 and F-value=231). Accordingly, we claim that the
logistic model can fit our data very well. Besides, we also have
studied the residuals of each empirical data and found that the
residuals are larger for the middle-class countries, say in the
second stage. Therefore, the export diversity can be predicted by
the logistic function better in the first and third stages.

\subsection{Three stages}
\label{sec.three}

From Fig \ref{fig.diversity_gdp}, we know that the curve can be
divided into three stages, namely initial stage, acceleration stage,
and final stage. However, there is no any unified and rigorous
method to identify them. In this paper, we use the three order
derivative of the logistic function (Equation \ref{eqn.scurve}) to
determine the stages accordingly,
\begin{equation}\label{eqn.deviate}
d^3(\frac{A}{1+e^{-k(x-X_M)}}) / dx^3 =0.
\end{equation}
Solving this equation, we can obtain two non-zero solutions, namely
the left cut-off point $(x_L,y_L)$ and the right cut-off point
$(x_R,y_R)$. We know that the left cut-off point which can separate
the initial stage and accelerating stage should maximize the
curvature of the curve. Therefore, setting the third order
derivative equaling zero can obtain this separating
point\citep{wang_delimiting_2009}. The same method follows for the
right cut-off point separating the accelerating stage and the final
stage.

There is another important point called inflection point $(X_M,Y_M)$
whose curvature is zero indicating the growth rate of the diversity
reaches the maximum. We can get the analytic expressions of these
important points in Table \ref{tab.points}.

\begin{table}
\centering
 \caption{The Coordinate of Three Important Points}
 {\small(We classify all
the sample countries into three stages according to them.)\\}
 \label{tab.points}
\begin{tabular}{ccccccc}
\hline  & Left Cut-off Point& Inflection Point& Right Cut-off Point\\
&$(X_L,Y_L)$  &$(X_M,Y_M)$ & $(X_R,Y_R)$ \\
\hline
X-coordinate&$X_M-\frac{\ln(2+\sqrt{3})}{k}$&$X_M$&$X_M-\frac{\ln(2-\sqrt{3})}{k}$&\\
&&&\\
Y-coordinate&$\frac{A}{3+\sqrt{3}}$&$\frac{A}{2}$&$\frac{A}{3-\sqrt{3}}$&\\
 \hline
\end{tabular}
\end{table}

Until now, the estimated curve in Fig \ref{fig.diversity_gdp} can be
divided into three stages by the two cut-off points. In the initial
stage, there are 71 countries including the poorest countries and
some developing countries such as Liberia, Nepal etc. Most of them
locate in Africa and Asia with less developed industries and
shortage of natural resources. The acceleration stage with 56
countries contains some developing and developed countries. Iran,
the second largest oil exporter, and Belarus, with developed
industry and advanced agriculture, are all developing countries and
locate below the inflection point. While, the developed countries,
such as New Zealand, Norway and Sweden, are at a high speed of
diversity growing with a falling acceleration above the inflection
point. Finally, the third stage contains the most developed
countries such as USA and Japan, with broad geographic and
well-developed industries. However, China, with GDP topped the ninth
in 1995, is also included in this stage. Guatemala and India both
are around the cut-off points. The former one will welcome to a
period of accelerating growth of export diversity while the later
one has experienced the glorious moments and will step into a period
of lower diversity growth rate.
\subsection{Dynamics}
\label{sec.alongwithtime}

\begin{figure}
\centerline {\includegraphics[width=18cm]{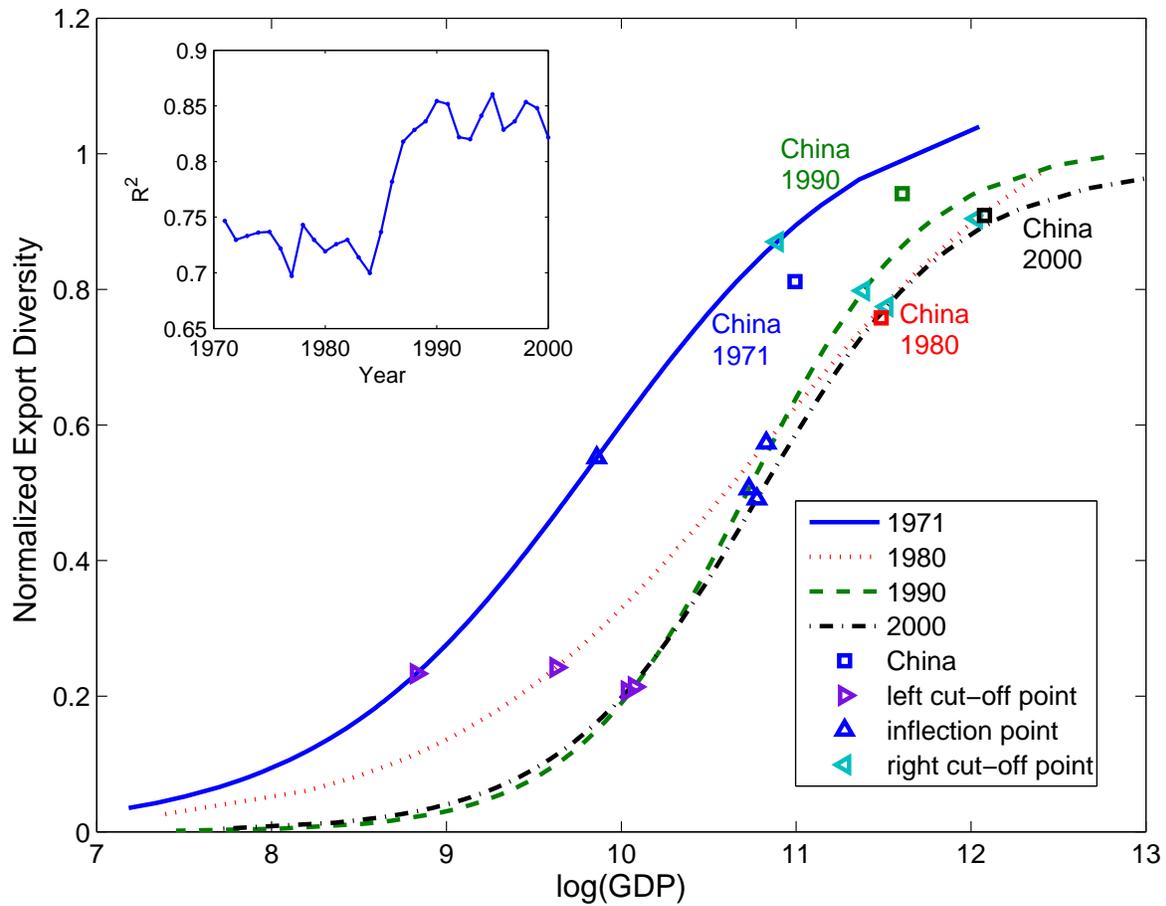}} \vskip3mm
\caption{The relationship between normalized diversity of export
goods and log(GDP) in four selected years. And the goodness of
fitting (R-square) in thirty years is shown in the
inset.}\label{fig.normalized_diversity}
\end{figure}

So far, we have studied the S-shaped pattern between export
diversity and log(GDP) only in one year. To look deeper into this
pattern especially on time, we fitted the collected data from 1971
to 2000 in the same way. However, because of the change of
classification method in 1983 mentioned in section \ref{sec.data},
we normalized the diversity data with the method below,
\begin{equation}\label{eqn.normalized}
Y_i'=\frac{Y_i-\min{\{Y_i\}}}{\max{\{Y_i\}}-\min{\{Y_i\}}},
\end{equation}
where $Y_i$ is the number of export goods categories of country $i$,
$Y_i'$ represents the normalized diversity which ranges from 0 to 1.
Fig \ref{fig.normalized_diversity} shows the patterns in four
different years with normalized data.

In Fig \ref{fig.normalized_diversity}, two cut-off points and the
inflection point are plotted in each year. We can see that the
S-shaped pattern is stable in these thirty years. The shift and the
expansion of the curves imply that the S-shaped pattern is more
obvious after 1990s. This phenomenon is also reflected by the
goodness of fitting in the inset of Fig
\ref{fig.normalized_diversity}. $R^2$ increases from 0.75 to around
0.83 in 1983, and fluctuates a little in the following years.

We can trace one specific country, China, which was undergoing
dramatic economic development in these years. We know that China has
adopted her opening policy from 1978, so she stayed in the
acceleration stage in 1971 and 1980 in Fig
\ref{fig.normalized_diversity}, meaning that the growth rate of the
diversity increased rapidly. After that, as her GDP ranked higher,
the growth rate of the diversity slowed down. Thus, China stepped
into the final stage.

\begin{figure} \centerline
{\includegraphics[width=15cm]{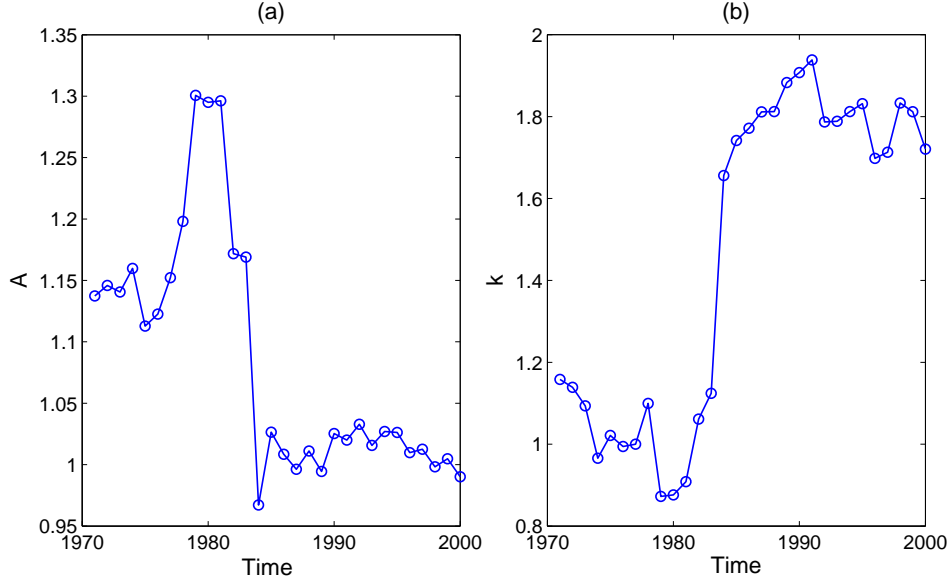}} \vskip3mm
\caption{The fluctuations of (a).$A$ (the upper asymptotes) and
(b).$k$ (the slope of the inflection point) in thirty
years}\label{fig.a_k_alongtime}
\end{figure}

\begin{figure}
\centerline {\includegraphics[width=15cm]{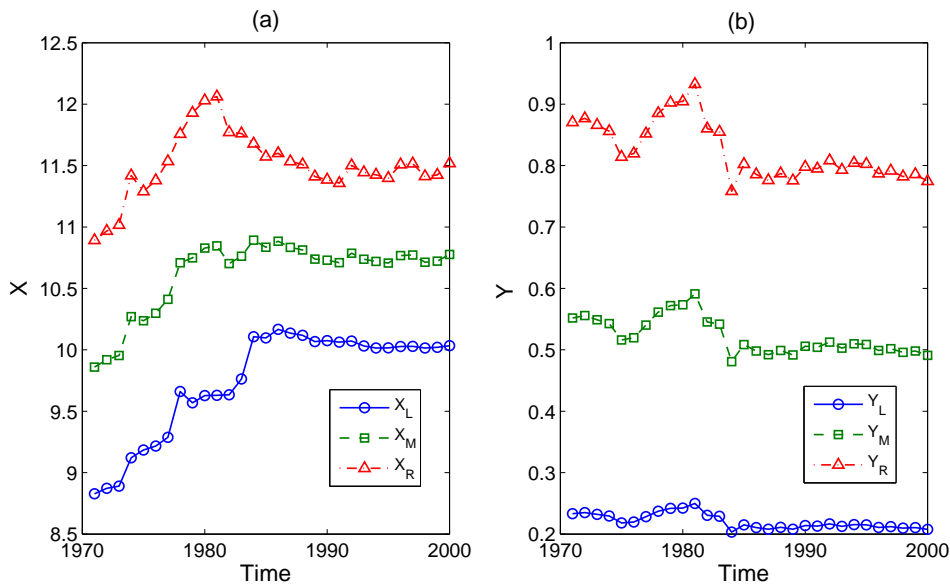}}
\vskip3mm \caption{The dynamical trends of the two cut-off points
and the inflection point during the 30 years (a). X-coordinates,
(b).Y-coordinates.}\label{fig.xandychange}
\end{figure}

From Fig \ref{fig.a_k_alongtime}(a) we know that $A$ fluctuates more
intensive before 1983 and then waves a little around 1. That means
the S-shaped relation is more stable after 1983. Fig
\ref{fig.a_k_alongtime}(b) shows the change of $k$ which is the
relative maximum growth rate of export diversity in the thirty
years. We know the value of $k$ keeps increasing from 1.13(1981) to
1.81(1990) while remains relatively stable from 1971 to 1979 and
1991 to 2000. Thus, the S-shaped curves became steeper suddenly in
about 1980-1985. This phenomenon can be also observed from the
dynamics of the three critical points. From Fig
\ref{fig.xandychange} we know the change of $X_L$, $X_M$ and $X_R$
values follow similar paths. In the first decade, the three
variables increase continually, and they remain 10.0, 10.7 and 11.5
respectively with only small fluctuations after 1983. However,
$Y_L$, $Y_M$ and $Y_R$ values do not have much change and remain
around their average value. Hence, the S-shaped curves move right
gradually at first and stop after 1980's. That indicates the global
economy kept increasing during the first decade and then slowed
down.

We believe that the dynamics of these parameters, especially the big
fluctuations around 1983, indicate the changed situation of the
international trade environment\citep{krugman_international_2002}.
First, we know that there was a serious global oil crisis in 1970s,
as a result, the system of international trade changed in 1980s all
around the world. Many countries abandoned the traditional
inward-looking trade, turned the export-substitution model of
economic growth to export-oriented model. Second, the adjustment of
the basic goal of market-oriented economic structure had been made
because of the boom of information and service industries in 1980s.
Furthermore, in order to adapt for the changes of the trade
environment, many countries made some change in 1980s. For instance,
the United State of America changed their trade policy to
multilateral free trade system due to the oil crisis and
depreciation. Japan, as another example, also strengthened the
multilateral trading system and developed the global economy since
1980s.

Thus, when the environment of the international trade turns better,
all countries tend to export more goods. This can be reflected by
the maximum growth rate of export diversity, $k$, which can indicate
the international trading circumstance. The environment changes
better as it jumps to a higher level.

\begin{figure}
\centerline
{\includegraphics[width=15cm]{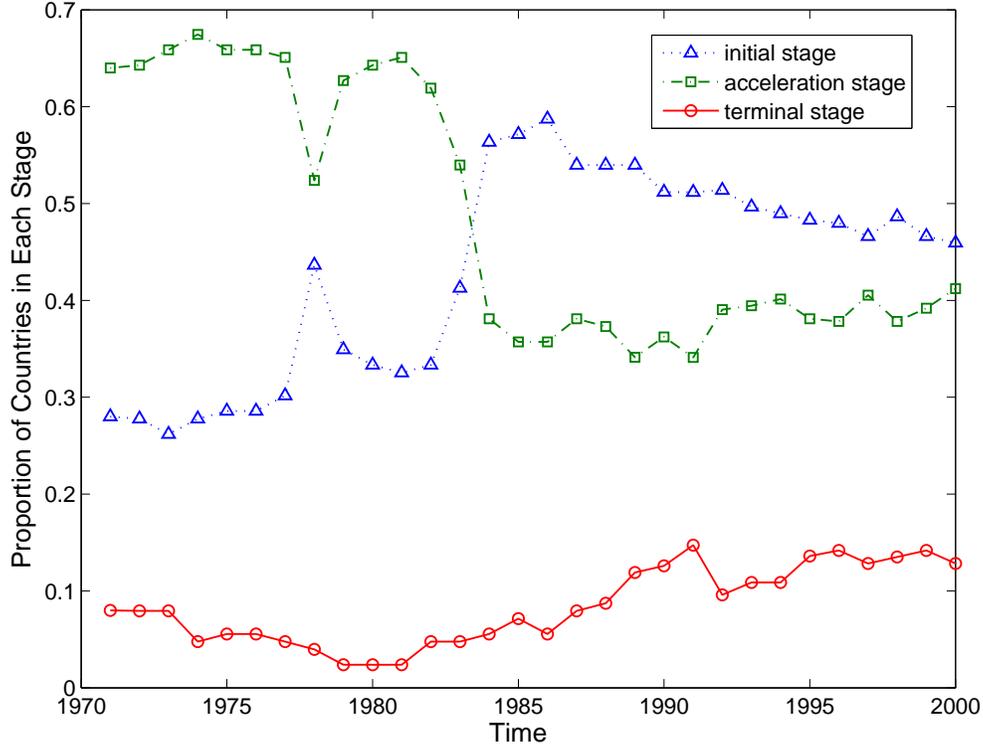}}
\vskip3mm \caption{The proportions of countries in each stage change
with time during the 30 years}\label{fig.proportions}
\end{figure}

Because $k$ increased and the total number of countries in the world
almost unchanged in these years, we know that the proportion of the
countries in each stage must have changed. From Fig
\ref{fig.proportions} we know that the proportion of the countries
in the initial stage increased while those in the acceleration stage
fell down during 1980 to 1985. After 1985, both fluctuated little
and the difference between them became smaller. However, the
proportion in the final stage almost remained around 0.1 all the
time. This phenomenon may be accounted for the unbalanced
development in different countries in the reform of international
trade policies as mentioned before. As the globalization process
started, the gap of the export diversity between the rich and poor
countries became larger. Consequently, the countries which had
climbed to the acceleration stage fell down into the initial stage
again.

\section{Discussion}
\label{rec.discussion} In previous sections, we have discussed the
S-shaped curves of the relationship between export diversity and log
(GDP). However, we have also the data of the other variables
including the number of importers, the categories of import goods,
and the number of exporters. So, we can explore more empirical
patterns of diversity. In this section, we will point out that all
the bivariate relationships fall into two classes, namely, S-shaped
relationships and power law relationships.

\subsection{S-shaped relationships}
\label{sec.sshapedrelationships}

\begin{figure}
\centerline {\includegraphics[width=15cm]{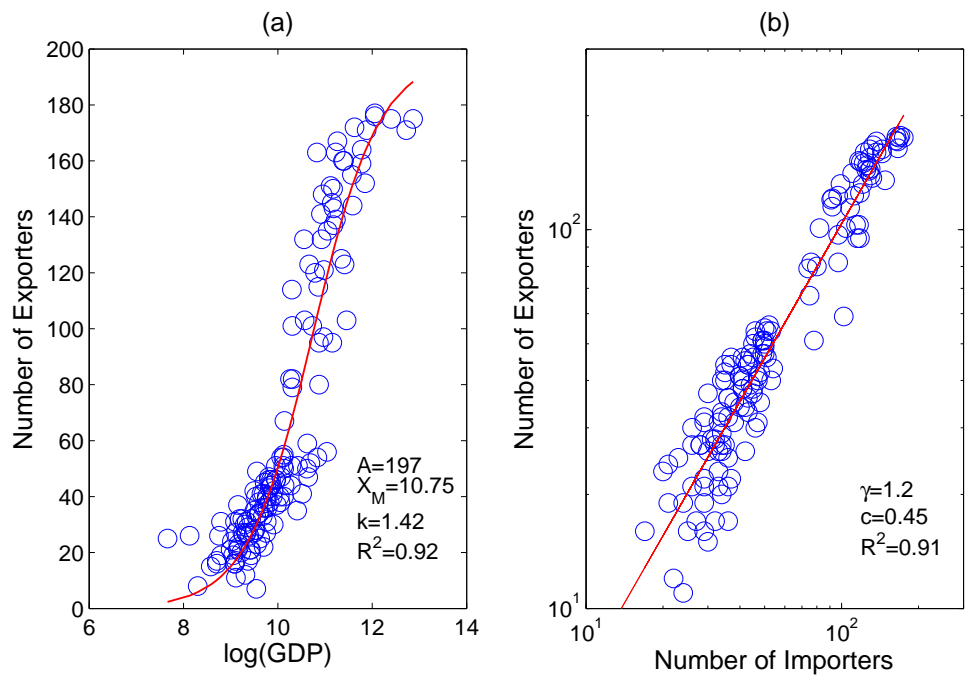}}
\vskip3mm \caption{Other relationships about diversity in 1995. (a).
The S-shaped curve of the relationship between the number of
exporters and the logarithm of GDP. (b). The power law relationship
between the number of importers and exporters.}
\label{fig.otherrelationship}
\end{figure}

We find that other diversity-related variables including the
categories of import goods, the number of distinct exporters and the
number of distinct importers are all have S-shaped relationship with
the logarithmic GDP for each country. Fig
\ref{fig.otherrelationship}(a) shows the relationship between the
number of exporters and log(GDP) as an example, and all these ``S''
curves are concluded in Table \ref{tab.srelationships}.

\begin{table}
\centering
 \caption{All the S-shaped Relationships}
 \label{tab.srelationships}
\begin{tabular}{ccccccc}
\hline  $x$& $y$& $A$&$k$&$X_M$&$R^2$\\
\hline
log(GDP)  & Importers& 203 &1.09&10.94&0.89\\
log(GDP)  & Exporters& 197 &1.42&10.75&0.92\\
log(GDP)  & Import Goods& 749 &1.66&9.65&0.87\\
log(GDP)  & Export Goods& 903 &1.85&10.7&0.86\\
\hline
\end{tabular}
\end{table}
As shown in Table \ref{tab.srelationships}, we know that the
logistic function can fit all the relationships between log (GDP)
and importers, exporters, import goods and export goods with high
significances ($R^2$). Nevertheless, the parameters ($A, k$ and
$X_M$ ) are very different which can also be divided into two
groups. We know that $A$ is the upper limit and represents the range
of diversity of trade. From Table \ref{tab.srelationships}, we
observe that $A_{import}<A_{export}$ for importers(exporters) but
$A_{import}>A_{export}$ for import(export) goods. That means
countries are more likely to export fewer kinds of goods to more
diverse countries. And $k$ is the maximum specific growth of $y$. So
$k_{import} <k_{export}$ always holds for countries and goods, which
means the export diversity grows faster than import diversity as GDP
grows.

\subsection{Power-law relationships}
\label{sec.powerlaw}

Scaling property is one of the most important universal quantitative
laws governing different dynamics of complex systems
\citep{brock_scaling_1999,brown_scaling_2000,lee_allometric_1989,jerome_chave_scale_2003,isalgue_scaling_2007,bettencourt_growth_2007,zhang_allometric_2010}.
In paper\citep{zhang_allometric_2010}, the authors have shown that
the import value and export value of a country all have power law
relationships with respect to GDP. In this paper, we find the
variables relating to diversity also follow the power law
relationship:
\begin{equation}\label{eqn.allometricscaling}
Y=cX^{\gamma},
\end{equation}
where, $X$ and $Y$ are diversity-related variables such as the
number of importers, categories of export goods etc. $c$ and
$\gamma$ are parameters to be estimated. By using equation
\ref{eqn.allometricscaling} to fit our empirical data, we obtain Fig
\ref{fig.otherrelationship}(b) and Table
\ref{tab.allometricscaling}.
\begin{table}
\centering
 \caption{All the Power Law Relationships}
 \label{tab.allometricscaling}
\begin{tabular}{ccccccc}
  \hline
  $X$&$Y$&$c$&$\gamma$&$R^2$\\
  \hline
  Importers & Exporters & 0.45& 1.2 & 0.91 \\
  Importers & Export Goods & 0.15& 1.75 & 0.78 \\
  Import Goods & Exporters & 0.052& 1.14 & 0.71 \\
  Import Goods & Export Goods & 0.0015& 1.92 & 0.79 \\
  Importers & Import Goods & 21.32 & 0.74 & 0.66 \\
  Exporters & Export Goods & 0.51 & 1.46 & 0.84 \\
  \hline
\end{tabular}
\end{table}

Fig \ref{fig.otherrelationship}(b)  shows the relationship between
the number of importers and exporters as an example. From Table
\ref{tab.allometricscaling}, we know that all power law
relationships are significant because their $R^2$s are all larger
than 0.6. And all the relationships between import properties and
export properties are super-linear\citep{bettencourt_growth_2007}
for $\gamma>1$, which means that with the increase of import
properties, export properties can grow faster. So do the
relationship between exporters and categories of export goods, that
is to say one country may have more exporters for each export goods
as the export goods diversity increases. However, the relationship
between importers and import goods is
sub-linear\citep{bettencourt_growth_2007} meaning that the growth
speed of the categories of import goods is slower than the number of
importers.

\section{Concluding Remarks}
\label{sec.conclusion}

We have shown the S-shaped relationship between the number of
different kinds of export commodities and log GDP, which is well
fitted by the logistic function. We have also found the two cut-off
points which can divide the curve into three stages. In addition, we
analyzed the dynamics of the parameters. Particularly, the parameter
$k$ increases suddenly in around 1980's which is supposed to
indicate the change of international trade environment. Finally, we
also discussed the other possible diversity-related properties all
have S-shaped relationships with respect to log GDP and power law
relations with other variables.

Although this paper did not give any explanations about the origins
of S-shaped curves and power law relationships, we believe that we
have made a first step to understand the economic diversity
phenomenon because we have revealed robust and universal patterns
based on the empirical data.

\paragraph{Acknowledgement}
We acknowledge the web site www.nber.org/data to provide free data
about international trade. This research was supported by National
Natural Science Foundation of China under Grants of No. 61004107


{\textbf{References}} 


\bibliographystyle{elsarticle-num}
\bibliography{diversity}







\end{document}